\documentclass[floatfix,pra,twocolumn,10pt]{revtex4}
\usepackage{amsmath}
\usepackage{amssymb}
\usepackage{times}
\usepackage{latexsym}
\usepackage{graphicx}
\usepackage[english]{babel}
\usepackage{verbatim}
\usepackage{bbm}
\usepackage[usenames,dvipsnames]{color}
\usepackage{hyperref}
\usepackage{tikz}

\newcommand{\eqs}{Eqs.~}
\newcommand{\fig}{Fig.~}

\newcommand{\ug} {\!=\!}

\newcommand{\piu} {\!+\!}
\newcommand{\meno} {\!-\!}

\newcommand{\rref} {Ref.~}
\newcommand{\rrefs} {Refs.~}
\usetikzlibrary{decorations.pathmorphing}

\begin{document}
\title{Steady-state entanglement activation in optomechanical cavities}
\author{Alessandro Farace$^{1}$, Francesco Ciccarello$^{2}$, Rosario Fazio$^{1}$ and Vittorio Giovannetti$^{1}$}
\affiliation{$^{1}$ NEST, Scuola Normale Superiore and Istituto Nanoscienze-CNR, I-56127 Pisa, Italy\\
$^{2}$ NEST, Istituto Nanoscienze-CNR and Dipartimento di Fisica e Chimica, Universit$\grave{a}$  degli Studi di Palermo, via Archirafi 36, I-90123 Palermo, Italy}
\date{\today}
%
%
%
%
%
%
\begin{abstract}
Quantum discord, and related indicators, are raising a relentless interest as a novel paradigm of non-classical correlations beyond entanglement. Here, we discover a  discord-activated mechanism yielding steady-state entanglement production
in a realistic continuous-variable setup. This comprises two coupled optomechanical cavities, where the optical modes (OMs) communicate through a fiber. We first use a simplified model to highlight the creation of steady-state discord between the OMs. We show next that such discord
improves the level of stationary optomechanical entanglement attainable in the system, making it more robust against temperature and thermal noise.
\end{abstract}
\maketitle
%
%
%
%

\section{Introduction}
\label{Sec:Int}
Entanglement arguably embodies the point where our classical-physics-based intuition conflicts the most with quantum mechanics. While abundance of experimental evidence has made this concept eventually accepted, recent work has shown that entanglement is not the only form of non-classical correlations. A composite system can happen to be in certain {\it mixed} states which, despite being unentangled,  feature correlations yet classically unexplainable [{quantum correlations} (QCs) in short]. Following the introduction of the so called {quantum discord} (QD)~\cite{Henderson1, Ollivier1}, a burst of attention to this new notion of non-classicality has arisen \cite{Modi1}. A major motivation comes from the fact that QD is the key resource enabling certain quantum information processing (QIP) schemes -- where entanglement is absent -- to outperform classical algorithms, see e.g.~\cite{BENEFIT1,datta,zeilinger,ralph,Madhok1,Hoban1,Rieffel1,Datta1}.

Unlike entanglement, production of discord-like QCs is not demanding since they can be created from classically-correlated states via local noise~\cite{Ciccarello1, streltsov,Ciccarello2, madsen,blatt}, a situation forbidding any entanglement to arise. In this respect, a rather spectacular effect that might have profound technological developments is {\it entanglement activation} 
(EA) via discord~\cite{Piani1,BRUSS,Piani2,Mazzola1} in a four-partite system. In short, this is the possibility to exploit the QCs between two (out of four) subparts -- yet fully disentangled -- in order to create entanglement across a bipartition of the global system. Arguably, this is possible because in non-classical states some amount of local quantum coherence is present, which can act as an entanglement-production catalyzer.

 \begin{figure}[t]
	\begin{center}
	\includegraphics[trim=0pt 0pt 0pt 0pt, clip, width=0.45\textwidth]{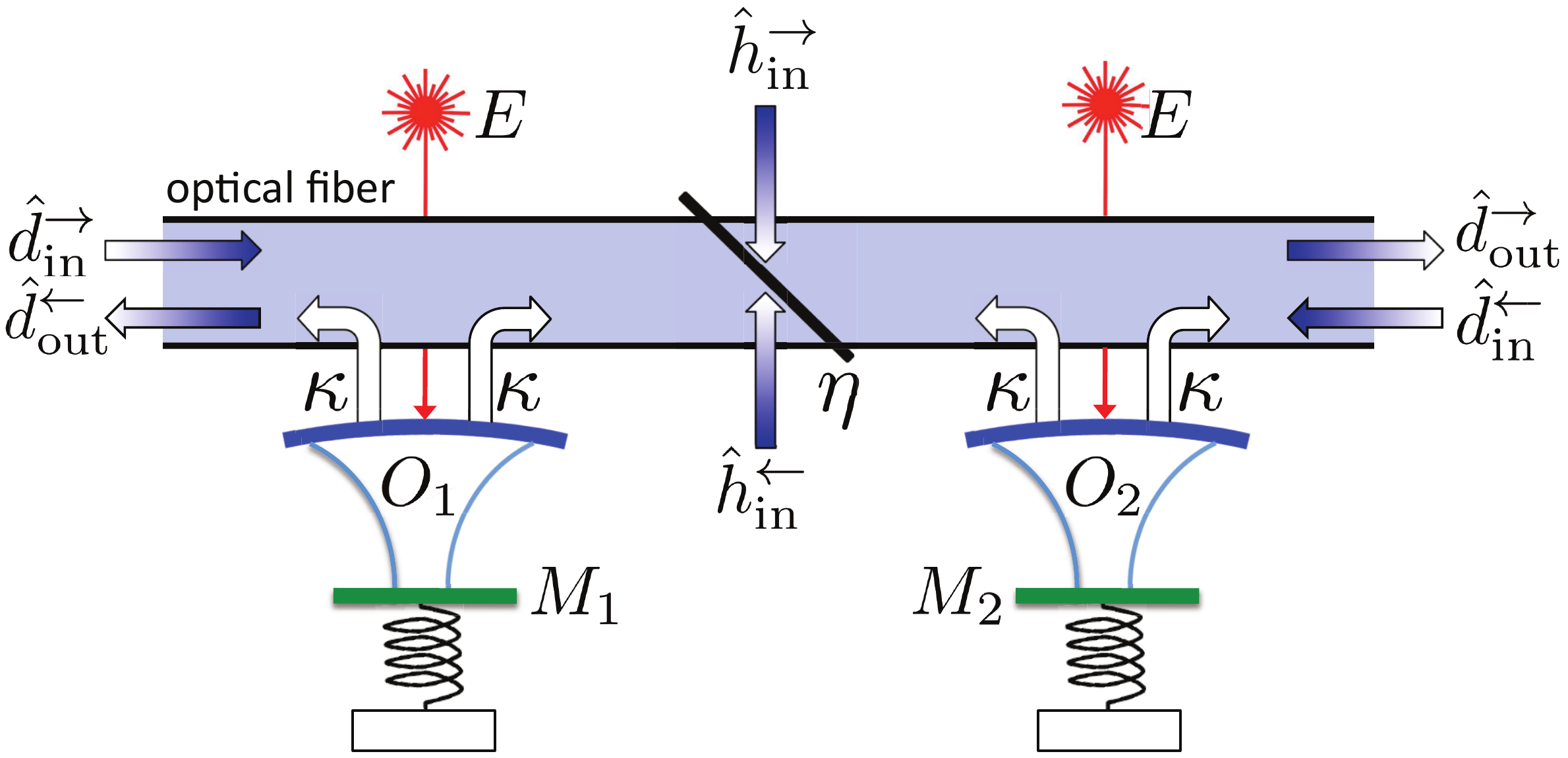}
  
		\caption{(Color online) Sketch of the system: two laser-driven optomechanical cavities coupled to an optical fiber. }		
		\label{FigCoupledOMs1}
	\end{center}
\end{figure}

Here, we show that it is possible to harness EA for improving steady-state entanglement-generation capabilities in realistic noisy settings,
 starting from a resource (non-classicality) which - at least for continuous variable (CV) systems - is easy to produce (essentially all bipartite Gaussian states are non-classically correlated \cite{Adesso1}).  
Opto-mechanical setups ~\cite{Aspelmeyer1, Genes1, Milburn1} are an ideal candidate for our investigation, given that entanglement production in such systems is currently a major challenge.
 Specifically, we discover a discord-activated mechanism allowing not only to increase but also to {\it maintain steadily} bipartite entanglement in a realistic optomechanical setup. Besides its fundamental relevance, this is clearly a paramount issue in view of a foreseeable technological exploitation of the
 EA mechanism and, furthermore, it well complies with
the spirit of the emerging paradigm of dissipation-driven QIP \cite{cirac, Diehl1}.

EA was first envisaged in terms of successive unitaries and finite-dimensional systems in noise-free scenarios~\cite{Piani1,BRUSS,Piani2}. It was recently extended to CV systems by Mazzola and Paternostro~\cite{Mazzola1}, who devised an attractive EA scheme in a pair of optomechanical cavities.
So far, though, only dynamical EA was demonstrated: The goal was to ensure that, at some instant of the considered evolution, entanglement is generated, no matter if this eventually fades away due to noise. 
Furthermore  in \rref\cite{Mazzola1} 
 the discord resource used
 for the enhancement generation   
  stems in fact from a two-mode photon entangled source, namely pre-existing entanglement is converted into QCs which are {\it afterwards} used for EA. Quite differently, besides producing a stationary entangled throughput, the mechanism we will present does not employ any entanglement supply in the input (in this specific respect, it can thus be regarded as a more genuine implementation of EA via discord).
  
We illustrate our findings in two steps.  First, in Sec. \ref{Sec:Disc}, we show a process yielding a steady-state amount of QCs between two cavity optical modes (OMs), where the employed resources are just two classically-correlated  input sources. This is achieved through fiber-mediated photon exchange between the OMs, each mode being additionally subject to a local noise source. Second, in Sec. \ref{Sec:Ent}, we consider two optomechanical cavities, where each OM is coupled to a noisy mechanical mode (MM) via radiation pressure (the local noises on the MMs being independent). Also, the two OMs can still exchange photons as in the previous step. We show that in this configuration, irrespective of the initial state of the MMs, the interplay between the optical discord production process and the radiation pressure activates entanglement across the optical-mechanical partition. 
As a pivotal feature, this entanglement persists indefinitely once steady conditions are reached, if and only if one keeps the coupling between the OMs (hence introducing discord). Conclusions follow in Sec. \ref{Sec:Conc}.

\section{Stationary throughput of quantum discord}
\label{Sec:Disc}
The setup we consider is sketched in Fig.~\ref{FigCoupledOMs1}. It comprises two identical optomechanical cavities 1 and 2, each made out of a single optical mode $O_j$ ($j\ug1,2$) interacting via radiation pressure with a corresponding single mechanical mode $M_j$ (see \rrefs \cite{Aspelmeyer1, Genes1, Milburn1} for a review on optomechanical systems). The two cavities are coupled to a common optical fiber, which enables the $O_1\!-\!O_2$ crosstalk crucial for the establishment of stationary QCs. The efficiency of this communication channel is measured by the fiber transmissivity $\eta$ with $0\!\le\!\eta\!\le\!1$ (see \fig1).
To illustrate the essentials of the QCs creation mechanism, in this section we use a {\it simplified model} where the pair of MMs is replaced by two independent thermal noise sources, which emulate the disturbance on the optical modes due to the radiation pressure coupling. For the sake of argument, we assume these optical noises to be fed via the input ports of the optical fiber as shown in \fig1. For now, each laser in \fig 1 can be neglected since a local displacement of the field operators cannot change the level of QCs.
Adopting the standard input-output formalism to tackle cascaded networks \cite{Gardiner1, Carmichael1}, the dynamics of $O_1$ and $O_2$ is described by a set of Langevin-type equations for their respective annihilation operators $\hat{a}_1$ and $\hat{a}_2$. These read
\begin{align}
\dot {\hat{a}}_1(t) \!&=\! - i \omega_C \hat{a}_1(t) \!-\! \kappa \hat{a}_1(t) \!-\! \kappa \sqrt{\eta}  
 \hat{a}_2(t-d/c) \nonumber \\\!&-\! \sqrt{\kappa}\left[  \hat d^{\rightarrow}_{\rm in} (t) \!+\!\sqrt{\eta}  \hat d^{\leftarrow}_{\rm in} (t\!-\!d/c)  
\!+\! \sqrt{1-\eta} \hat h^{\leftarrow}_{\rm in} (t)\right]\!\! \nonumber \\
\dot {\hat{a}}_2(t) \!&=\! - i \omega_C \hat{a}_2(t) \!-\! \kappa \hat{a}_2(t) \!-\! \kappa \sqrt{\eta}  
 \hat{a}_1(t-d/c) \nonumber \\\!&-\! \sqrt{\kappa}\left[  \hat d^{\leftarrow}_{\rm in} (t) \!+\!\sqrt{\eta}  \hat d^{\rightarrow}_{\rm in} (t\!-\!d/c)  
\!+\! \sqrt{1-\eta} \hat h^{\rightarrow}_{\rm in} (t)\right]\!\!, \label{EqCoupledOscillatorsLangevin}
\end{align}
where the two cavity modes have identical frequency $\omega_C$ and linewidth $\kappa$, and $d/c$ is the time taken by the output signals to travel the inter-cavity distance (see App. \ref{AppChannel} for a detailed derivation). Without loss of generality, we set $d/c\ug0$ henceforth. Noise fluctuations are described by four independent bath annihilation operators $\hat d^{\rightarrow}_{\rm in} (t)$, $\hat d^{\leftarrow}_{\rm in} (t)$, $\hat h^{\rightarrow}_{\rm in} (t)$, and $\hat h^{\leftarrow}_{\rm in} (t)$, each fulfilling white-noise commutation rules, i.e. $[\hat d^{\rightarrow}_{\rm in} (t), \hat d^{\rightarrow\dag}_{\rm in} (t')]\ug \delta(t\meno t')$ and analogous identities. The superscript arrows specify the direction of propagation of the associated degree of freedom along the fiber (see Fig.~\ref{FigCoupledOMs1}). In particular, $\hat d^{\rightarrow}_{\rm in} (t)$ and ${\hat {d}^{\leftarrow}}_{\rm in} (t)$ describe the two independent thermal sources which, in this simplified picture, emulate the effect of the MMs.  
Their temperature is set through the identities $\langle \hat{d}^{\rightarrow\dag}_{\rm in} (t) \hat d^{\rightarrow}_{\rm in}(t') \rangle \ug 
\langle d^{\leftarrow\dag}_{\rm in} (t) \hat d^{\leftarrow}_{\rm in}(t') \rangle \ug n_{\rm in} \delta(t-t')$, where $\langle\cdots \rangle$ is the expectation value over the bath input state and $n_{\rm in}$  is the bath mean photon number. $\hat h^{\rightarrow}_{\rm in} (t)$ and $\hat h^{\leftarrow}_{\rm in} (t)$ are the vacuum noise operators associated with the loss along the fiber and fulfill $\langle \hat h^{\rightarrow\dag}_{\rm in} (t) \hat h^{\rightarrow}_{\rm in}(t') \rangle = 
\langle \hat h^{\leftarrow\dag}_{\rm in}(t) \hat h^{\leftarrow}_{\rm in}(t') \rangle =0$.

Eqs.~\eqref{EqCoupledOscillatorsLangevin} show two
possible mechanisms that can establish QCs: An effective direct coupling between $\hat a_1$ and $\hat a_2$ and, in addition, the correlation between the total noise on $O_1$ and that on $O_2$.
To quantify the QCs between the continuous-variable systems $O_1$ and $O_2$, we adopt the Gaussian discord ${\cal D}_G$~\cite{Adesso1, Giorda1}, a measure (see App. \ref{AppDiscord} for details) that can be used in the present problem due to the linearity of Eqs.~\eqref{EqCoupledOscillatorsLangevin} and the Gaussian nature of the input noises (the asymptotic state of the system is thereby Gaussian too).

\begin{figure}[t]
	\begin{center}
	\includegraphics[trim=0pt 0pt 0pt 0pt, clip, width=0.35\textwidth]{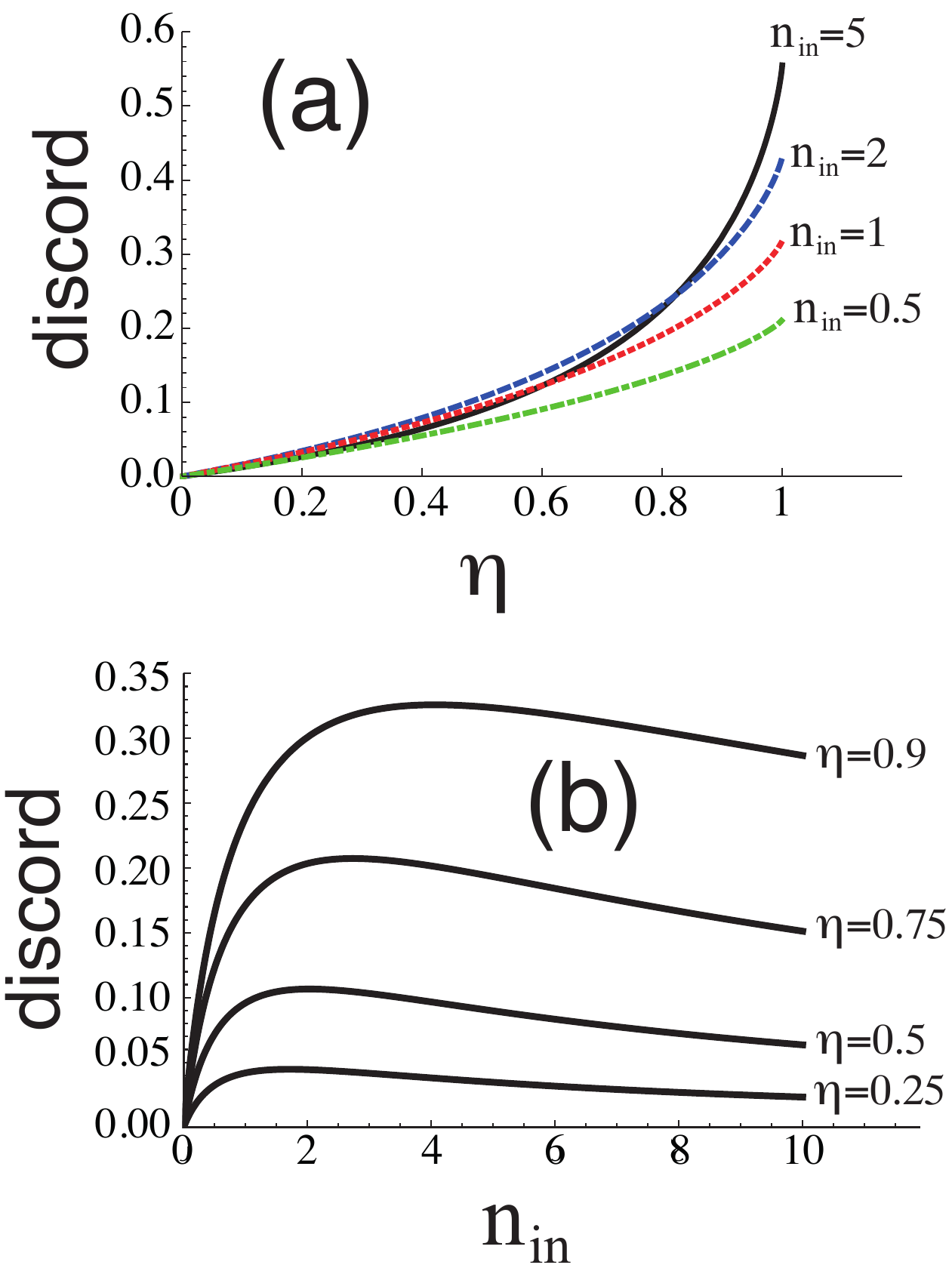}
\vspace{5pt}
		\caption{(Color online) Asymptotic Gaussian discord ${\cal D}_G$ between $O_1$ and $O_2$ against $\eta$ (a) and $n_{\rm in}$ (b) for different values of $n_{\rm in}$ and $\eta$, respectively, as predicted by Eqs.~(\ref{EqCoupledOscillatorsLangevin}). The plots are independent of the values taken by $\omega_C$ and $\kappa$.}
		\label{FigCoupledOscillatorsDiscord}
	\end{center}
\end{figure}

In Fig.~\ref{FigCoupledOscillatorsDiscord}, we study the dependance of the asymptotic value of ${\cal D}_G$ on $n_{\rm in}$ and the fiber transmittivity $\eta$. Evidently, any non-zero value of $\eta$ always yields a finite amount of QCs  ($O_1$ and $O_2$ are fully independent when $\eta=0$, hence QCs cannot arise).
In particular, as shown by \fig2(a), the discord monotonically increases with the transmissivity  $\eta$ of the waveguide (hence with the intensity of the coupling between the two modes)~\cite{NOTA1}. 
Also, note that discord is created provided that the reservoirs associated with $\hat d^{\rightarrow}_{\rm in} (t)$ and $\hat d^{\leftarrow}_{\rm in} (t)$ are at non-zero temperature [see \fig2(b)], namely $n_{\rm in}\!\neq\!0$. Indeed, if the temperature is zero the asymptotic cavity state is the vacuum featuring no correlations at all. On the other hand, $\mathcal D_G$ asymptotically vanishes for high $n_{\rm in}$ since 
at high temperatures decoherence is too strong for QCs to arise. Thereby, discord is a non-monotonic function of the bath temperature. As for entanglement between $O_1$ and $O_2$ instead, this identically {\it vanishes} regardless of $\eta$ and $n_{\rm in}$, as can be checked by computing the logarithmic negativity (see App. \ref{AppEntanglement} for details on this measure of entanglement). Hence, as a key feature of our mechanism, the fiber-mediated link between the cavities is unable to entangle $O_1$ and $O_2$ but, as shown, can establish significant discord between them.

\section{Entanglement activation}
\label{Sec:Ent} 
Next, to show the usefulness of the discord creation mechanism discussed so far, we consider the full optomechanical system in \fig1 and prove that an EA mechanism can take place. The MMs' degrees of freedom now enter the dynamics explicitly. In the proper rotating frame, the Hamiltonian of the $j$th optomechanical cavity thus reads
 \begin{eqnarray}
\hat H_{j}\ug   - \Delta_0 \hat{a}^\dag_j   \hat{a}_j   \piu   \omega_M \dfrac{\hat{q}_j^2 \piu \hat{p}_j^2}{2} \meno G_0 \hat{a}^\dag_j   \hat{a}_j   \hat{q}_j  \piu i   E (\hat{a}^\dag_j   
\meno\hat{a}_j   )\,,
\end{eqnarray}
where  $\hat{q}_j$ and $\hat{p}_j$ are the canonical coordinates  of $M_j$ with
$\omega_M$ being the associated frequency, $G_0$ is the optomechanical coupling strength, while $E$ is the coupling rate to an external driving laser of frequency $\omega_C \piu\Delta_0$ (the detuning $\Delta_0$ is assumed to be small compared to $\omega_C$). Including the interaction with the environment in a way analogous to the previous section,
we end up with a set of coupled quantum Langevin-type equations (this time involving {\it both} the optical {\it and} the mechanical degrees of freedom). These read
\begin{eqnarray} 
\left\{ \begin{array}{l}
\dot{\hat{q}}_j = \omega_M  \hat{p}_j,\\
\dot {\hat{p}}_j = -\omega_M  \hat{q}_j - \gamma  \hat{p}_j + G_0 \hat{a}_j  ^\dag \hat{a}_j  
  +  \hat \xi_j,\\
\dot {\hat{a}}_1 = i \Delta_0  \hat{a}_1 + i G_0  \hat q_1  \hat{a}_1  + E - k  \hat{a}_1  - k \sqrt{\eta}  \hat{a}_2   \\
\qquad \quad  - \sqrt{k}\left[  \hat d^{\rightarrow}_{\rm in}  +\sqrt{\eta}  \hat d^{\leftarrow}_{\rm in} 
+ \sqrt{1 - \eta} \,\hat h^{\leftarrow}_{\rm in} \right]\;,  \\ 
\dot  {\hat{a}}_2 =i \Delta_0  \hat{a}_2 + i G_0 \hat q_2 \hat{a}_2   + E - k  \hat{a}_2  - k \sqrt{\eta}  \hat{a}_1  \\ 
\qquad \quad  - \sqrt{k}  \left[   \hat d^{\leftarrow}_{\rm in} +\sqrt{\eta}  \hat d^{\rightarrow}_{\rm in} + \sqrt{1 - \eta} \,\hat h^{\rightarrow}_{\rm in} \right]\;.
\end{array}
\right.
\label{EqOptomechanicsCoupled}
\end{eqnarray}
Here, $\gamma$ is the damping rate of each MM while $\hat \xi_{j}(t)$ stands for the associated Gaussian noise operator fulfilling white noise commutation relations. $\hat \xi_{1}(t)$ and $\hat \xi_{2}(t)$ are independent but have the same temperature, set through the identity $\langle \hat \xi_j(t) \hat \xi_{j'}(t') \rangle \ug \gamma (2n_{\rm M} \piu 1) \delta(t \meno t') \delta_{jj'}$ with $n_{\rm M}$ being the thermal excitation number of the mirror fluctuations. All the remaining parameters and operators have the same meaning as in Eqs.~(\ref{EqCoupledOscillatorsLangevin}). Differently from the simplified model discussed earlier, however,  we now set to zero the mean photon number of $\hat d^{\leftarrow}_{\rm in}$ and $\hat d^{\rightarrow}_{\rm in}$  (i.e,. $n_{\rm in}=0$) as there is no longer need for `emulating' the MMs~\cite{NOTANEW1}.

The essential parameters that we use to obtain our findings are $\omega_M/2\pi = 947$ KHz, $\gamma_M/2 \pi = 140$ Hz, $\Delta_0 = -\omega_M$, $k/2\pi = 215$ KHz, $G_0 = 24$ Hz, $E=4\times10^{11}$ (corresponding to a laser power of $11$ mW). These match the realistic setup in \rref\cite{REALISTIC}. 
In particular, we assume red-detuned (i.e., $\Delta_0\!<\!0$) and intense lasers being shined on the system in a way that $E$ is strong enough to achieve ground-state cooling of the MMs \cite{Genes1}. In this regime, we can approximate Eqs.~\eqref{EqOptomechanicsCoupled} as a set of classical equations for the mean values $\{\langle \hat{q}_j \rangle$,$\langle  \hat{p}_j \rangle$, $\langle \hat{a}_j   \rangle\}$ and a set of linearized equations for the corresponding quantum fluctuations $\{\delta \hat{q}_j=\hat{q}_j - \langle \hat{q}_j\rangle$, $\delta \hat{p}_j= \hat{p}_j-\langle \hat{p}_j \rangle$, $\delta  \hat{a}_j  =\hat{a}_j  - \langle \hat{a}_j  \rangle\}$.
As all the noise operators are Gaussian, the system dynamics and its steady state are fully specified once the first and second momenta of the field operators are known.  

\begin{figure}[t]
	\begin{center}
	\includegraphics[trim=0pt 0pt 0pt 0pt, clip, width=0.4\textwidth]{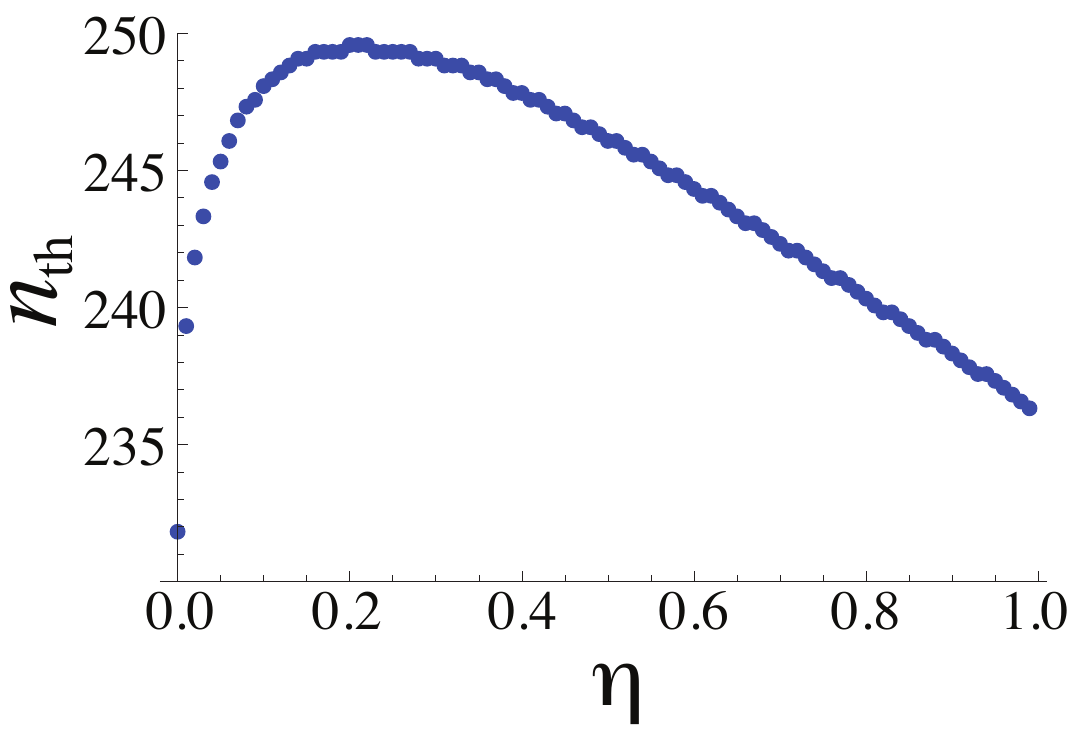}
		\caption{(Color online) Threshold value of $n_{\rm M}$ for the appearance of stationary entanglement between $M_1M_2$ and $O_1O_2$ (as quantified by $\mathcal E$) against the fiber transmitivity $\eta$. }
		\label{soglia}
	\end{center}
\end{figure}
\begin{figure}[t]
	\begin{center}
	\includegraphics[trim=0pt 0pt 0pt 0pt, clip, width=0.35\textwidth]{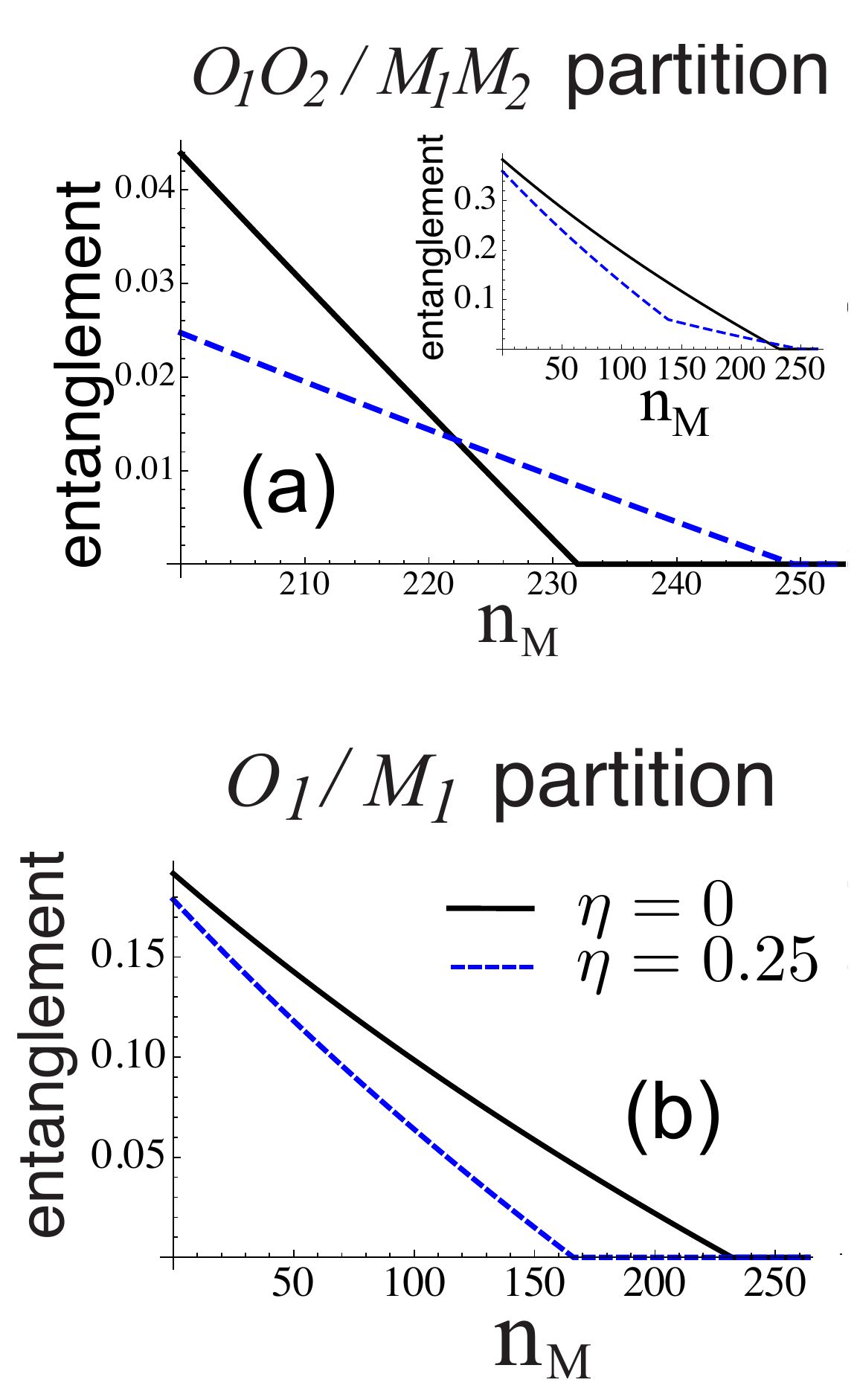}

		\caption{(Color online) Stationary entanglement (as measured by the LN $\mathcal E$) associated with the bipartition  $O_1 O_2/M_1 M_2$ (a) and  $O_1 /M_1$ (b) against the thermal excitation number $n_{\rm M}$ for $\eta=0$ (black solid line) and $\eta=0.25$ (blue dashed). In panel (a), a narrow interval of $n_{\rm M}$ is displayed (region across the crossing point) in order to highlight the central effect of our mechanism. Inset: full behavior of $O_1 O_2/M_1 M_2$ entanglement.}
		\label{fig4}
	\end{center}
\end{figure}

We will analyze the amount of optomechanical entanglement in order to assess whether it benefits from the presence of discord. To measure the entanglement between the MMs and OMs, we use the logarithmic negativity (LN) $\mathcal E$~\cite{Vidal1} associated with  
the $O_1O_2/M_1M_2$ bipartition, which is a suitable measure of entanglement for Gaussian states. We point out that, when $\eta\ug0$, $\mathcal E$ exactly quantifies this entanglement since in this case the two optomechanical cavities are independent and all the $O_1O_2/M_1M_2$ correlations reduce to the two-mode $O_j$/$M_j$ correlations.
When  $\eta \!>\! 0$ instead, $\mathcal E$ yields only a lower bound for the $O_1O_2/M_1M_2$ entanglement since this measure is not faithful when genuine 4-mode correlations are involved \cite{Werner1}. In particular, a null value of $\mathcal E$ does not imply the absence of optomechanical entanglement (more details on this can be found in App. \ref{AppEntanglement}).

As is known, provided that the temperature is below a threshold value of $n_{\rm M}$, which we will call $n_{\rm th}$, steady-state optical-mechanical entanglement can be created \cite{Vitali1}. In Fig.~\ref{soglia}, we plot the threshold temperature associated with the stationary $O_1O_2/M_1M_2$  entanglement as a function of $\eta$. Remarkably, the presence of the fiber raises $n_{\rm th}$ for any value of $\eta\!>\!0$. 
 In particular, while for $\eta\ug0$ entanglement survives up to temperatures of the order of $n_{\rm th} \!\sim\! 230$, for $\eta\simeq 0.25$ this becomes as high as $n_{\rm th}\! \sim\! 250$ with an enhancement of almost 10$\%$. 
In  \fig4(a), we compare the stationary LN across the $O_1O_2/M_1M_2$ bipartition as a function of $n_{\rm M}$ for $\eta\ug0$ with  $\eta\ug0.25$. The fiber clearly brings about a two-slope behavior in such a way that $\mathcal E$ is lowered for values of $n_{\rm M}$ up to $n_{\rm M}\!\sim\! 220$ but enhanced beyond this point. This results in an improved tolerance of entanglement to thermal noise [see region on the right of the crossing point in \fig4(a)]. We show next that such additional entanglement is of a genuine multipartite nature and clarify the mechanism responsible for its formation.

Different values of $\eta$ yield different solutions of \eqs \eqref{EqOptomechanicsCoupled} at the classical level, hence the equations for the operators' fluctuations depend on different strengths of the effective optomechanical coupling $G = G_0 \left< \hat a_j \right>$. This  fact  alone could, in principle, increase the entanglement between $O_j$ and $M_j$ ($j\ug1,2$) without building any crossed correlations. 
To show that this is not the case, in Fig.~4(b) we study the stationary LN between one OM and its mechanical counterpart (say $O_1$ and $M_1$). Notably, this specific entanglement is {\it always} reduced by a finite transmissivity $\eta\!>\!0$ (namely, in the presence of the fiber) compared to the $\eta\ug0$ case. The joint occurrence of this behavior and the entanglement enhancement with respect to the $O_1O_2/M_1M_2$ bipartition in \fig 4(a) thus provides evidence that {\it crossed} correlations between the optical and mechanical parts \cite{Akram1} are necessarily built up during the dynamical evolution (see also App. \ref{AppActivation} for details). 

To highlight the role of discord in the augmented entanglement 
production, we next 
focus on the dynamics of the system in the transient time. In particular in \fig5, we compare the time behavior of $\mathcal E$ across the $O_1O_2/M_1M_2$ bipartition with that of the Gaussian discord between $O_1$ and $O_2$, having set the temperature of the mechanical baths to $n_{\rm M}\ug240$ and the coupling to $\eta=0.25$. Hence, in the light of \fig4 (a), we are in a regime where the fiber-mediated coupling, as signaled by $\eta\! \neq \!0$, is crucial for the generation of entanglement. The OMs (MMs) are initially prepared in the vacuum state (thermal state with mean occupation number $n_{\rm M}=240$).
The system develops a non-zero  ${\cal D}_G$  (black line) which, in line with the simplified model of the previous section, after a transient, stabilizes around an asymptotic value ($\sim 0.0139$ for the specific parameters we used). Concomitantly, $O_1O_2/M_1M_2$ entanglement also arises (blue line) and reaches a steady value, but only after some discord is present in the system. While
 a direct comparison between the
  values of  ${\cal E}$ and ${\cal D}_G$ is not possible (the two measures being 
 both unbounded and not convertible into each other), the plot   provides a clear evidence that discord is needed in order for entanglement to appear. Indeed, we remark  that the steady state $O_1O_2/M_1M_2$ entanglement is always accompanied by a steady-state discord between $O_1$ and $O_2$. On the contrary, if the coupling is absent ($\eta=0$) there is no entanglement at $n_{\rm M}=240$, but also no discord can be produced since the two cavities are completely independent. A detailed discussion on the functional dependence  of ${\cal E}$ and ${\cal D}_G$ upon the
system parameters can be found in appendix \ref{AppActivation}.

Importantly, still in line with the behavior of the simplified model, the entanglement between $O_1$ and $O_2$ is identically {\it zero} for all values of $\eta$.
\fig5 is hence the first theoretical evidence of an entanglement-activation mechanism
producing a {\it stationary} throughput of multipartite entanglement between four modes (two mechanical and two optical), without extracting it from pre-existing entanglement sources. 

\begin{figure}[t]
	\begin{center}
	\includegraphics[trim=0pt 0pt 0pt 0pt, clip, width=0.4\textwidth]{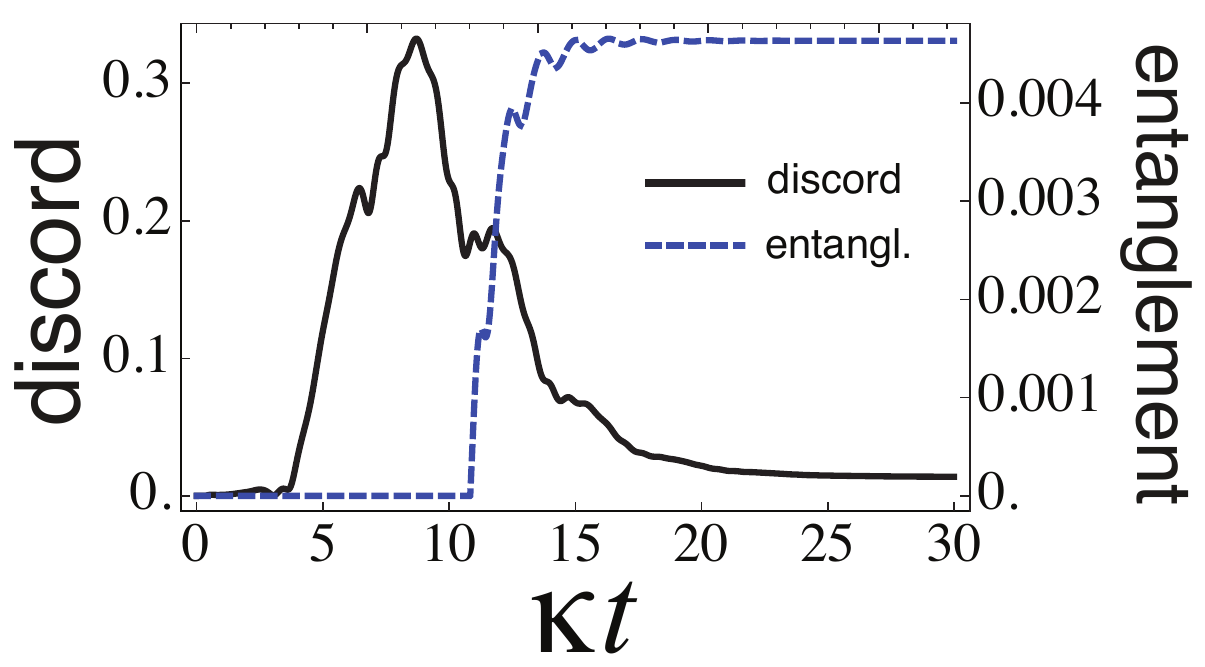}
		\caption{(Color online) Time evolution of the Gaussian discord ${\cal D}_G$ between $O_1$ and $O_2$ (black continuous line) and of the logarithmic negativity $\mathcal E$ 
		across the $O_1 O_2/M_1 M_2$ bipartition (blue dashed line).  We set $n_{\rm M} = 240$, $\eta=0.25$ and taken the OMs initially in the vacuum state. Time is measured in units of $\kappa^{-1}$.}
		\label{fig6}
	\end{center}
\end{figure}\mbox{}

\section{Conclusions}
\label{Sec:Conc}
We showed an EA scheme via quantum discord in two optomechanical cavities, where the OMs interact through a fiber. The fiber-enabled crosstalk creates significant discord between the OMs, while leaving them fully disentangled. Such mechanism affects stationary entanglement across the optical-mechanical bipartition so that it survives at temperatures for which it would not be seen without the fiber. Remarkably, such discord-activated entanglement is of a genuinely multipartite nature.

Recent developments in the fabrication of optomechanical crystals \cite{Eichenfield1} allow for the on-chip realization of both photonic and phononic waveguides, together with localized optical and mechanical resonances \cite{Safavi1, Safavi2}. The co-localization of mechanical and optical resonances enables high values of optomechanical coupling. Moreover, the possibility of evanescent coupling between the localized resonances and the waveguides has been proven \cite{Safavi2}. Optomechanical crystals thereby appear a promising scenario for a not-far-fetched experimental implementation of our scheme. Quantitatively similar results can indeed be found with parameters matching the typical scales of optomechanical crystals.
Also, the capabilities offered by these systems make it interesting to look at different scenarios. This may include coupling the mechanical modes to a common reservoir, or coupling near localized resonances (both optical and mechanical) in a coherent way (by photon or phonon tunneling). 

\section{Acknowledgments}
\label{Sec:Ack}
We thank M. Aspelmeyer, M. Paternostro and T. Tufarelli for comments and discussions. This work was supported by the EU projects SIQS and NANOCTM, by the MIUR-PRIN
(``Collective quantum phenomena: From strongly
correlated systems to quantum simulators")
 and by the MIUR-FIRB-IDEAS project RBID08B3FM.
%
%
%
%

\appendix
\section{Derivation of the Langevin equations for the optical modes}
\label{AppChannel}
We show here that the waveguide coupled to the two cavities can be described in terms of two unidirectional channels. We follow the original derivation made by Gardiner \cite{Gardiner1} for a single unidirectional channel. We have a 1-dimensional electromagnetic bath (the waveguide) which couples to cavity $1$ at position $x=0$ and with cavity $2$ at position $x=d$. The system-bath Hamiltonian can be written as
\begin{eqnarray}
H &=& H_{S} + \int_{-\infty}^{\infty} \! dk \; \hbar c |k| 	\hat d^\dagger(k) \hat d(k) \\
&&+ i \hbar \int_{-\infty}^{\infty} \! dk \; \Gamma_1(k) \left\{ \hat a_1 \hat d^\dagger(k) - \hat a_1^\dagger \hat d(k) \right\} \nonumber \\
&&+ i \hbar \int_{-\infty}^{\infty} \! dk \; \Gamma_2(k) \left\{ \hat a_2 \hat d^\dagger(k) e^{-i k d} - \hat a_2^\dagger \hat d(k) e^{i k d} \right\}. \nonumber 
\end{eqnarray}
$\hat a_1$ and $\hat a_2$ are destruction operators for the radiation modes of cavity $1$ and $2$ (both have frequency $\omega_C$). $\hat d(k)$ is the destruction operator associated with the bath mode of wavevector $k$. $\Gamma_1(k)$ and $\Gamma_2(k)$ are the coupling strengths of cavity $1$ and $2$ with the bath. We write the Heisenberg equation for $\hat d(k)$
\begin{equation}
\dot {\hat d}(k) = - i c |k| \hat d(k) + \Gamma_1(k) \hat a_1 + \Gamma_2(k) \hat a_2 e^{-ikd}
\end{equation}
which can be formally solved to give
\begin{eqnarray}
&& \hat d(k,t) = e^{- i c |k| (t-t_0)} \hat d(k,t_0) + \int_{t_0}^{t} \! ds \; e^{- i c |k| (t-s)} \Gamma_1(k) \hat a_1(s) \nonumber \\
&&\quad + \int_{t_0}^{t} \! ds \; e^{- i c |k| (t-s)} \Gamma_2(k) \hat a_2(s) e^{-ikd}.
\label{EqFormalD}
\end{eqnarray}
We substitute eq \eqref{EqFormalD} into the Heisenberg equation for $\hat a_1$, writing the $k>0$ modes and the $k<0$ modes separately.
\begin{equation}
\dot {\hat a}_1 = - i \omega_C \hat a_1 - \int_{-\infty}^{\infty} \! dk \;  \Gamma_1(k) \hat d(k,t)
\end{equation}
\begin{widetext}
\begin{gather}
\dot {\hat a}_1 = - i \omega_C \hat a_1 - \int_{k>0}^{} \! dk \;  \Gamma_1(k) e^{- i c k (t-t_0)} \hat d_0(k) - \int_{k<0}^{} \! dk \;  \Gamma_1(k) e^{i c k (t-t_0)} \hat d_0(k) \nonumber \\
- \int_{k>0}^{} \! dk \;  \Gamma^2_1(k) \int_{t_0}^{t} \! ds \; e^{- i c k (t-s)} \hat a_1(s) - \int_{k<0}^{} \! dk \;  \Gamma^2_1(k) \int_{t_0}^{t} \! ds \; e^{i c k (t-s)} \hat a_1(s) \nonumber \\
- \int_{k>0}^{} \! dk \;  \Gamma_1(k) \Gamma_2(k) \int_{t_0}^{t} \! ds \; e^{- i c k (t-s)} \hat a_2(s) e^{-ikd} - \int_{k<0}^{} \! dk \;  \Gamma_1(k) \Gamma_2(k) \int_{t_0}^{t} \! ds \; e^{i c k (t-s)} \hat a_2(s) e^{-ikd}.
\label{EqFormalA}
\end{gather}
\end{widetext}
We assume that all the coupling is within a narrow range of $c |k| \sim \omega_C$, and constant in this range:\linebreak $\Gamma_1(\pm \omega_C/c) = \sqrt{\kappa_1/2 \pi}$, $\Gamma_2(\pm \omega_C/c) = \sqrt{\kappa_2/2 \pi}$. (In the main text we further set $\kappa_1=\kappa_2=\kappa$ for simplicity.) We can then make the approximations \cite{Gardiner2}
\begin{gather}
\int_{k>0}^{} \! dk \;  \Gamma_{1,2}(k) e^{- i c k t} = \sqrt{2 \pi \kappa_{1,2}} \delta(t),
\end{gather}
\begin{gather}
\int_{k<0}^{} \! dk \;  \Gamma_{1,2}(k) e^{i c k t} = \sqrt{2 \pi \kappa_{1,2}} \delta(t).
\end{gather}
Eq \eqref{EqFormalA} becomes
\begin{eqnarray}
&& \dot {\hat a}_1 = - i \omega_C \hat a_1 - \int_{k>0}^{} \! dk \;  \Gamma_1(k) e^{- i c k (t-t_0)} \hat d_0(k) \nonumber \\
&& - \int_{k<0}^{} \! dk \;  \Gamma_1(k) e^{i c k (t-t_0)} \hat d_0(k) \nonumber \\
&& - \kappa_1 \hat a_1(t) - \sqrt{\kappa_1 \kappa_2} \hat a_2(t-d/c).
\end{eqnarray}\\
The same derivation can be done for $\hat a_2$ and we have
\begin{eqnarray}
&& \dot {\hat a}_2 = - i \omega_C \hat a_2 - \int_{k>0}^{} \! dk \;  \Gamma_2(k) e^{- i c k (t-t_0)} e^{i k d} \hat d_0(k) \nonumber \\
&&- \int_{k<0}^{} \! dk \;  \Gamma_2(k) e^{i c k (t-t_0)} e^{ikd} \hat d_0(k)  \nonumber \\ 
&& - \kappa_2 \hat a_2(t) - \sqrt{\kappa_1 \kappa_2} \hat a_1(t-d/c).
\end{eqnarray}\\
In both equations, we can identify the first integral (over $k>0$ modes) as an input field going from left to right and the second integral (over $k<0$ modes) as an input field going from right to left. Defining \cite{Gardiner3}
\begin{gather}
\sqrt{\frac{1}{2\pi}} \int_{k>0}^{} \! dk \;  e^{- i c k (t-t_0)} \hat d_0(k) \equiv \hat d^{\rightarrow}_{in} (t),
\end{gather}
\begin{gather}
\sqrt{\frac{1}{2\pi}} \int_{k<0}^{} \! dk \;  e^{i c k (t-t_0)} e^{i k d} \hat d_0(k) \equiv \hat d^{\leftarrow}_{in} (t),
\end{gather}
we finally get
\begin{widetext}
\begin{gather}
\dot {\hat a}_1 = - i \omega_C \hat a_1 - \sqrt{\kappa_1} \hat d^{\rightarrow}_{in} (t) - \sqrt{\kappa_1} \hat d^{\leftarrow}_{in} (t-d/c) - \kappa_1 \hat a_1(t) - \sqrt{\kappa_1 \kappa_2} \hat a_2(t-d/c),
\end{gather}
\begin{gather}
\dot {\hat a}_2 = - i \omega_C \hat a_2 - \sqrt{\kappa_2} \hat d^{\rightarrow}_{in} (t-d/c) - \sqrt{\kappa_2} \hat d^{\leftarrow}_{in} (t) - \kappa_2 \hat a_2(t) - \sqrt{\kappa_1 \kappa_2} \hat a_1(t-d/c).
\end{gather}
\end{widetext}
This is formally equivalent to having two separate unidirectional channels, which redirect the output of one cavity to the other. For example if we take the $\rightarrow$ channel, the output from the left cavity $\hat d^{\rightarrow}_{out} (t) = \hat d^{\rightarrow}_{in} (t) + \sqrt{\kappa_1} \hat a_1 (t)$ plays as an additional input for the right cavity (with some delay). The opposite happens in the $\leftarrow$ channel.

For the sake of realism, we can also introduce losses along the waveguide. We model them by inserting a beam-splitter located somewhere between the two cavities, which couples the guided modes to the vacuum outside. The beam-splitter has transmittivity $\eta$. In this way, the right cavity sees the bare input $d^{\leftarrow}_{in} (t)$ plus the output of the left cavity mixed with a vacuum noise $\hat h^{\rightarrow}_{in} (t)$, i.e. $\sqrt{\eta} \hat d^{\rightarrow}_{out} (t-d/c) + \sqrt{1-\eta} \hat h^{\rightarrow}_{in} (t)= \sqrt{\eta} \hat d^{\rightarrow}_{in} (t-d/c) + \sqrt{\eta} \sqrt{\kappa_1} \hat a_1 (t-d/c) + \sqrt{1-\eta} \hat h^{\rightarrow}_{in} (t)$. Final equations are then
\begin{widetext}
\begin{gather}
\dot {\hat a}_1 = - i \omega_C \hat a_1 - \sqrt{\kappa_1} \hat d^{\rightarrow}_{in} (t) - \sqrt{\eta} \sqrt{\kappa_1} \hat d^{\leftarrow}_{in} (t-d/c) - \kappa_1 \hat a_1(t) - \sqrt{\eta} \sqrt{\kappa_1 \kappa_2} \hat a_2(t-d/c) - \sqrt{1-\eta}\sqrt{\kappa_1} \hat h^{\leftarrow}_{in} (t),
\end{gather}
\begin{gather}
\dot {\hat a}_2 = - i \omega_C \hat a_2 - \sqrt{\eta} \sqrt{\kappa_2} \hat d^{\rightarrow}_{in} (t-d/c) - \sqrt{\kappa_2} \hat d^{\leftarrow}_{in} (t) - \kappa_2 \hat a_2(t) - \sqrt{\eta} \sqrt{\kappa_1 \kappa_2} \hat a_1(t-d/c)  - \sqrt{1-\eta}\sqrt{\kappa_2} \hat h^{\rightarrow}_{in} (t).
\end{gather}
\end{widetext}
%
%
%
%
%
%
\section{Gaussian Discord}
\label{AppDiscord}

Quantum discord \cite{Henderson1, Ollivier1} has been recently proposed as measure of quantum correlations between two parties $A$ and $B$ which is more general than entanglement, e.g. there exist separable states with non-zero discord. By definition, quantum discord is the difference $\mathcal{D}(B|A)=\mathcal{I}(AB)-\mathcal{J}(B|A)$ between total correlations $\mathcal{I}(AB)$, as measured by quantum mutual information
\begin{equation} \label{equazionenuova}
\mathcal{I}(AB) = S(\rho_A) + S(\rho_B) - S(\rho_{AB}),
\end{equation}
and classical correlations $\mathcal{J}(B|A)$, interpreted as the information gain about one subsystem ($B$) as a result of a measurement on the other ($A$).
\begin{equation}
\mathcal{J}(B|A) = \max_{\{E_a\}} \left[ S(\rho_{B}) - \sum_a p_a S \left( \frac{Tr_A[\rho_{AB} E_a]}{p_a} \right) \right],
\end{equation}
where$S(\rho)$ is the Von Neumann entropy, $\sum_a E_a = \mathbb{I}$ is a positive-operator valued measure (POVM) on $A$ and $p_a = Tr[\rho_{AB} E_a]$ is the probability of outcome $a$.

Originally proposed for qubits, the concept has been generalized to gaussian states in continuous-variable systems \cite{Adesso1, Giorda1}, under the name of gaussian discord ${\cal D}_G$. This is obtained by restricting the
optimization in Eq.~(\ref{equazionenuova})  
to Gaussian POVM. As a consequence ${\cal D}_G$
provides in general only a lower bound for ${\cal D}$ 
(namely, states with non zero values of ${\cal D}_G$ 
will certainly exhibits a certain degree of discord).   
For Guassian states however it is conjectured to be optimal, i.e. ${\cal D}_G = {\cal D}$~\cite{Adesso1, Giorda1, GIORDA,OLIV}.
 An analytic form is known to compute gaussian discord for all possible two-mode gaussian states. Notably, all two-mode gaussian states, with the exception of product states ($\rho_{AB} = \rho_A \otimes \rho_B$), have finite gaussian discord.

In the main text, we are interested in the discord between two optical modes, so we report the explicit formula referring to the specific case of a two-mode gaussian state. We take the $4\times4$ correlation matrix $\mathcal{C}$
\begin{equation}
\mathcal{C}_{ij} = \frac{1}{2} \left< v_i v_j + v_j v_i \right>,
\end{equation}
where $\vec v = (\delta \hat x_1, \delta \hat y_1, \delta \hat x_2, \delta \hat y_2)^\top$ is the vector of quadratures' deviation from their mean value; i.e. 
\begin{equation}
\hat x_i = \frac{\hat a_i^\dagger + \hat a_i}{\sqrt{2}}, \;\;\;\;\; \hat y_i = i \frac{\hat a_i^\dagger - \hat a_i}{\sqrt{2}}, \;\;\;\;\; \delta \hat x_i = \hat x_i - \left< \hat x_i \right>.
\end{equation}
$\mathcal{C}$ can be written in the $(2\times2)$-blocks form
\begin{equation}
\mathcal{C} = \left( \begin{array}{cc} \mathcal{C}_1 & \mathcal{C}_3 \\ \mathcal{C}_3^\top & \mathcal{C}_2 \end{array} \right).
\end{equation}
From the correlation matrix $\mathcal{C}$, five symplectic invariants \cite{Ferraro1} can be constructed
\begin{eqnarray}
	&I_1=4 \;\mbox{Det}[\mathcal{C}_1], \;\;\; I_2=4\; \mbox{Det}[\mathcal{C}_2], \;\;\; I_3=4 \; \mbox{Det}[\mathcal{C}_3],&
	\nonumber \\
	& I_4=16\;  \mbox{Det}[\mathcal{C}], \;\;\; I_\Delta=I_1 + I_2 + 2I_3, \nonumber&
\end{eqnarray}
and two symplectic eigenvalues
\begin{gather}
	\lambda_{\pm}=\sqrt{\frac{I_\Delta \pm \sqrt{I_\Delta^2-4I_4}}{2}}.
	\label{EqSympEig}
\end{gather}
These quantities, which are invariant under local unitary operations, are the natural building blocks from which the measure of gaussian discord (also invariant under local unitaries) can be constructed.
\begin{equation}
	\mathcal{D}_G(B|A) = f(\sqrt{I_1})-f(\lambda_{-})-f(\lambda_{+})+f(\sqrt{W}),
\end{equation}
where
{\begin{equation}
	f(x) \equiv \left( \tfrac{x+1}{2} \right) \log \left( \tfrac{x+1}{2} \right) - \left( \tfrac{x-1}{2} \right) \log \left( \tfrac{x-1}{2} \right)
\end{equation}}
and 
\begin{widetext}
\begin{gather}
	W = \begin{cases}
		\dfrac{2I_3^2+(I_1-1)(I_4-I_2)+2|I_3|\sqrt{I_3^2+(I_1-1)(I_4-I_2)}}{(I_1-1)^2} & \;\;\; \text{if } (I_4 - I_2 I_1)^2 \leq (1+I_1)I_3^2(I_2+I_4)\\\\
		\dfrac{I_2 I_1 - I_3^2 + I_4 - \sqrt{I_3^4+(I_4-I_2 I_1)^2-2I_3^2(I_4+I_2 I_1)}}{2I_1} & \;\;\; \text{otherwise},
	\end{cases}
\end{gather}
\end{widetext}
(in the above equations and hereafter  the logarithm are expressed in base 2).
%
%
%
%
%
%

\section{Logarithmic negativity}
\label{AppEntanglement}

In the main text, we want to compute the entanglement for various bipartite ($1\otimes1$-modes or $2\otimes2$-modes) gaussian states of a continuous variable system. A convenient measure of entanglement for such states is the so-called logarithmic negativity. It directly stems from the positive partial transpose (PPT) criterion \cite{Peres1bis} for discriminating entangled and separable states. A bipartite separable state can be written by definition as {$\rho_{SEP} = \sum_i p_i \rho_A^{(i)} \otimes \rho_B^{(i)}$, with 
$\rho_A^{(i)}$, $\rho_B^{(i)}$ being states of the subsystems $A$ and $B$ respectively and $p_i$ being probabilities.} It's easy to see that its partial transpose with respect to one subsystem (say A) $\rho_{SEP}^{\top_A} = \sum_i p_i \rho_A^{(i)\top_A} \otimes \rho_B^{(i)}$ is still a valid density matrix and hence is positive definite. Conversely, a non positive partial transpose always indicates the presence of entanglement. The logarithmic negativity quantifies how negative the partial transpose is.

For $1\otimes1$-modes gaussian states the PPT criterion is both necessary and sufficient \cite{Simon1bis}. This also implies that the logarithmic negativity is a faithful measure of entanglement. We report the analytic formula and a sketch of its derivation, using the same notation of the previous section. At the level of correlation matrix $\mathcal{C}$, partial transposition is equivalent to changing the sign of momenta for a subsystem (say A). The partial transpose $\mathcal{C}^{\top_A}$ is positive if and only if its symplectic eigenvalue $\tilde \lambda_{-}$ is greater than $1/2$ \cite{Ferraro1}. The symplectic eigenvalue $\tilde \lambda_{-}$ can be found, analogously to eq \eqref{EqSympEig}, as
\begin{gather}
	\tilde \lambda_{-}=\sqrt{\frac{\tilde I_\Delta - \sqrt{\tilde I_\Delta^2-4I_4}}{2}},
\end{gather}
where now $\tilde I_\Delta=I_1 + I_2 - 2I_3$ (note the change of sign due to partial transposition). The logarithmic negativity $\mathcal{E}$ is then defined as
\begin{equation}
	\mathcal{E} = \max\{ 0, -\log(2 \tilde\lambda_{-}) \}.
\end{equation}
Consistently $\mathcal{E} > 0$ when $\tilde\lambda_{-} < 1/2$.

For a $2\otimes2$-modes gaussian system the picture is more complicated. The PPT criterion for separability becomes only necessary, i.e. $\rho_{SEP}^{\top_A} \ngeqslant 0 \Rightarrow entanglement$ but the opposite is not true in general \cite{Werner1}. The formula for the logarithmic negativity becomes more involved as well. First, the partial transposed matrix will have four symplectic eigenvalues $\tilde \lambda_j$: they can be computed as the eigenvalues of the matrix $| i  \Omega \mathcal{C}^{\top_A}|$, where $\Omega$ is the symplectic matrix
\begin{equation}
\Omega = \bigoplus_1^4 \left( \begin{array}{cc} 0 & 1 \\ -1 & 0 \end{array} \right).
\end{equation}
Second, multiple symplectic eigenvalues can be smaller than $1/2$ and we need to sum the various contributions. In the end 
\begin{equation}
	\mathcal{E} = \sum_j \max\{ 0, -\log(2 \tilde\lambda_{j}) \}.
\end{equation}
However, since the PPT criterion is not sufficient, we could still have an entangled state with $\mathcal{E} = 0$ (the measure is not faithful). The logarithmic negativity can be then considered as a lower bound for the entanglement in the system. 
It is also worth observing that as in the case of the Guassian discord ${\cal D}_G$, ${\cal E}$ can assume
arbitrarily high values.

\subsection{Relation between Gaussian discord and entanglement}

No direct connection between the values of ${\cal E}$ and ${\cal D}_G$ can be established. However for very large values of the entanglement (as measured by the Gaussian entanglement of formation) -- see e.g. \cite{Adesso1} --  it is known that the value of Gaussian discord becomes proportional to the value of the Gaussian entanglement of formation (EoF), hence a connection between the two is restored. Unfortunately, the results presented in the paper are in a regime of low entanglement, and choosing Gaussian EoF over LN brings no benefit to the analysis.
%
%
%
%
%
%

\section{Remark on the entanglement behavior}
\label{AppActivation}

For $\eta \neq 0$, the entanglement $O_1 O_2/M_1M_2$ between the optical part and the mechanical part has a discontinuous derivative when plotted against $n_M$, with a slow decaying tail which survives up to higher temperatures ($n_M \sim 250$). Most importantly, this behavior is peculiar to the bipartition $O_1 O_2/M_1M_2$. If we look at the optomechanical entanglement in any other bipartition (i.e. $O_1/M_1$, $O_2/M_1$, $O_1 O_2/M_1$ or $O_1/M_1 M_2$), the curve has a simple decay and reaches zero well below $n_M \sim 250$, without any sudden change in the slope.

\begin{figure}[t]
	\begin{center}
	\includegraphics[trim=0pt 0pt 0pt 0pt, clip, width=0.4\textwidth]{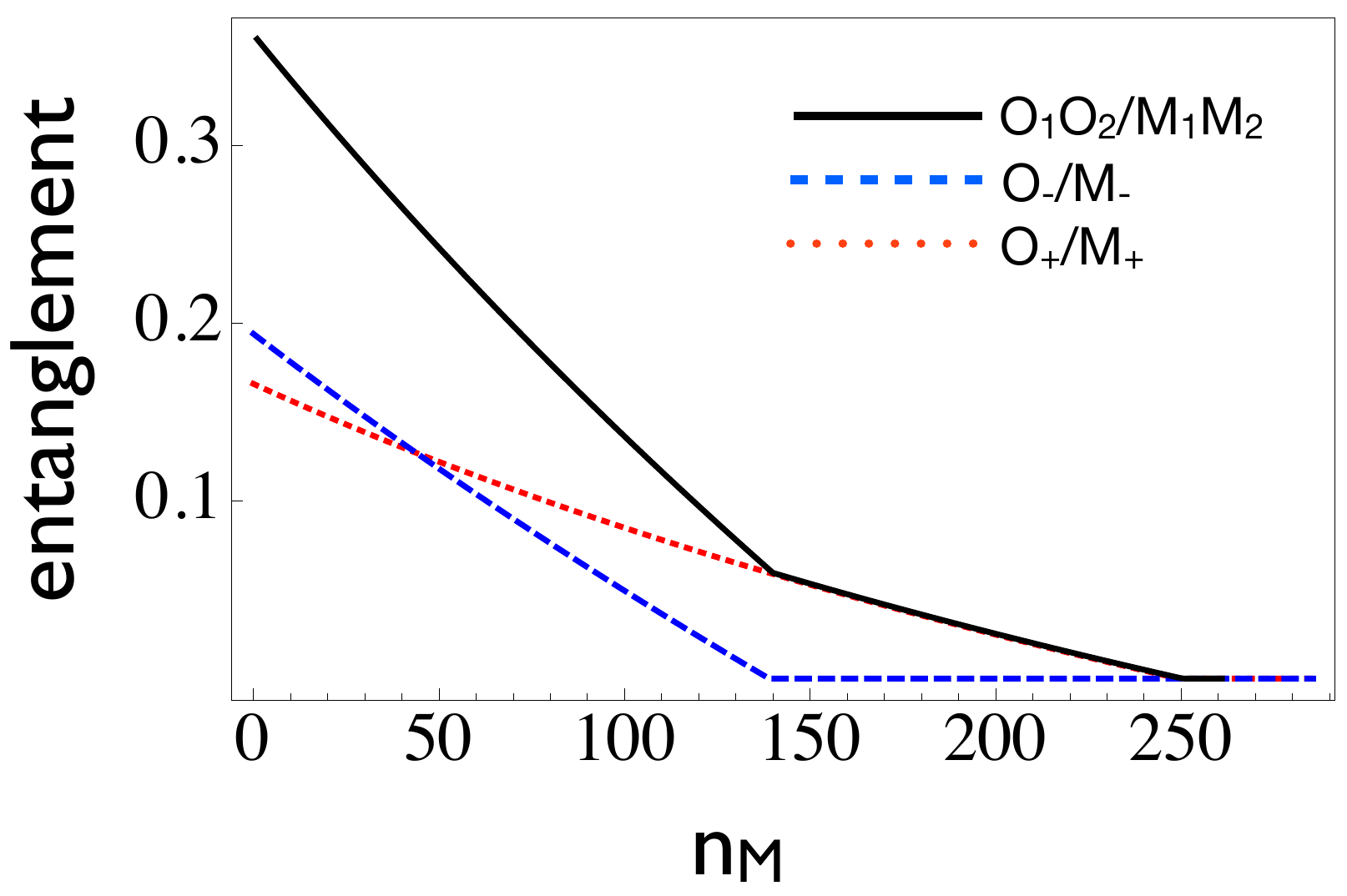}
		\caption{(Color online) Stationary entanglement (measured in terms of the logarithmic negativity) between the mechanical mode $M_+$ and the optical mode $O_+$ (red dotted line), for different values of the environment temperature (as measured by the mean number $n_M$ of thermal excitations in the mechanical mode). Stationary entanglement between the mechanical mode $M_-$ and the optical mode $O_-$ (blue dashed line). The sum of the two curves (black solid line) coincides with the stationary entanglement across the $O_1 O_2/ M_1 M_2$ bipartition. $\eta = 0.25$.}
		\label{FigCoupledEntanglementRotatedN}
	\end{center}
\end{figure}

We can deduce that the robust component of the $O_1 O_2/M_1M_2$ entanglement is given by correlations between some global combination of mechanical modes $\alpha q_1 + \beta q_2$ and some global combination of the optical modes $\gamma a_1 + \delta a_2$. Thanks to the symmetry of our specific setting, we guess that these modes are of the form $q_{\pm} = \tfrac{q_1 \pm q_2}{\sqrt{2}}$ and $x_{\pm}=\tfrac{x_1 \pm x_2}{\sqrt{2}}$ (where $x_j$ represents the quadrature $(a_j+ a_j^\dag)/\sqrt{2}$). By repeating the calculations in the new basis, we find that there is no entanglement between $q_{+}$ and $x_{-}$ (or between $q_{-}$ and $x_{+}$). The equations for the $+$ modes, are indeed decoupled from those of the $-$ modes. Entanglement is present between $q_{-}$ and $x_{-}$ but survives only up to $n_M \sim 150$. Entanglement between $q_{+}$ and $x_{+}$ survives instead up to $n_M \sim 250$, thus explaining the double-component nature of the $O_1 O_2/M_1M_2$ entanglement. This also shows that the increase in the entanglement is due to the presence of crossed correlations between the optomechanical systems 1 and 2.
Moreover, as seen from Fig.~\ref{FigCoupledEntanglementRotatedN}, we find that the sum of the two contributions 
gives precisely the total $O_1 O_2/M_1M_2$ entanglement found in the main text.

\subsection{Characterizing the entanglement production}
A precise, quantitative characterization 
 on the complex interplay which links the  logarithmic negativity  ${\cal E}$ 
of the  $O_1 O_2/M_1 M_2$ bipartition to
 the Gaussian discord ${\cal D}_G$ of the optical modes $O_1$ and $O_2$, is made difficult by the presence of
 several  parameters   which play a double role in the system dynamics (for instance the
bath temperature contribute both to compromise the entanglement
production and to EA mechanisms by providing the fuel needed
for the discord generation).  Yet some useful insights can be obtained by  studying how these parameters 
affects the temporal evolution of ${\cal E}$ and 
${\cal D}_G$.

\begin{figure}[h]
	\begin{center}
	\includegraphics[trim=0pt 0pt 0pt 0pt, clip, width=0.4\textwidth]{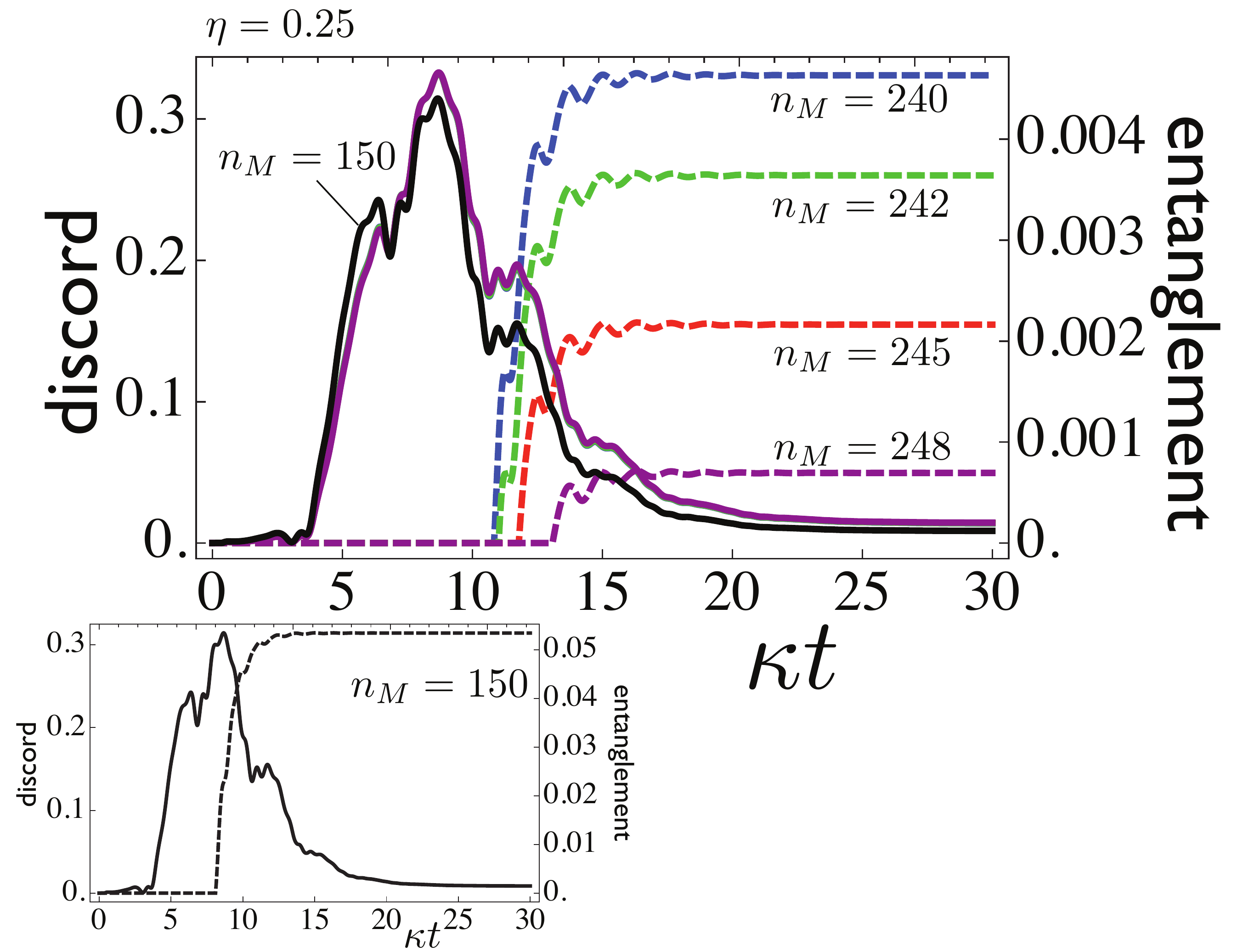}
		\caption{(Color online) Time evolution of the Gaussian discord ${\cal D}_G$ between $O_1$ and $O_2$ (continuous lines) and of the logarithmic negativity $\mathcal E$ 
		across the $O_1 O_2/M_1 M_2$ bipartition (dashed lines). Different color refer to different values of the thermal excitation number $n_M$. While the entanglement production  is 
highly sensitive to variation of $n_M$  (the higher $n_M$ the	smaller ${\cal E}$),  the temporal dependence of ${\cal D}_G$ appears  not to be affected by small variation of this parameter:
all the curves of ${\cal D}_G$ obtained for $n_M=248$ to $n_M=240$ are overlapping. To see significant variations in the temporal behavior of ${\cal D}_G$ one needs to reach $n_M=150$
(black curve). The corresponding value of entanglement is presented in the inset of the figure: due to the low values of the thermal noise	in this case the entanglement reaches values of ${\cal E}$ which are 10 time larger than those obtained for $n_M=240$.   
	In all plots	 we have set $\eta=0.25$ and taken OMs  to be initially in the vacuum state. Time is measured in units of $\kappa^{-1}$.}
		\label{Figsupp1}
	\end{center}
\end{figure}

Figure.~\ref{Figsupp1} illustrates the dependence of 
${\cal D}_G$ and ${\cal E}$ upon 
 the thermal excitation number $n_M$ (i.e.  the temperature of the oscillator bath). 
 As in the case of  Fig. 5 of the main the text, for all the values
 of $n_M$ we have considered, 
 both quantities 
reach stationary values after a transient time interval where
the maximum of  ${\cal D}_G$  is followed by a sharp increase of ${\cal E}$. As expected from the sensitivity of entanglement with respect
to noise, the plot shows that even small increases in  $n_M$  have
a rather detrimental effect on  ${\cal E}$: in particular the asymptotic  value of the latter is a decreasing function of
   $n_M$. 
 On the contrary ${\cal D}_G$ appears to be
 insensitive to small variations in $n_{M}$ 
 (all the curves associated with values of $n_M$ within few percent from $n_M=240$  overlap). Effects of the change
 of the bath temperature become evident only at lower
 values of $n_{M}$. An example is provided by the black
 continuous curve of the figure which represents the temporal
 evolution of ${\cal D}_G$ for $n_M=150$: the associated level of discord gets reduced with respect to the cases where $n_M\simeq 240$ (this is consistent with the fact that the bath temperature
is responsible for  triggering  the discord production in the model). Notice also that for such low value of $n_M$
 the entanglement generation is larger  by a factor of $10$ 
with respect to the level obtained for $n_M\simeq 240$ -- see inset. In this regime however the temperature of the system
is already low enough to ensure  that  the opto-mechanical coupling alone is capable to generate entanglement between in the individual 
opto-mechanical systems (i.e. $O_1/M_1$ and
$O_2/M_2$) without the aid of EA mechanism -- see Fig. 4(b) of the main text.

\begin{figure}[th]
	\begin{center}
	\includegraphics[trim=0pt 0pt 0pt 0pt, clip, width=0.4\textwidth]{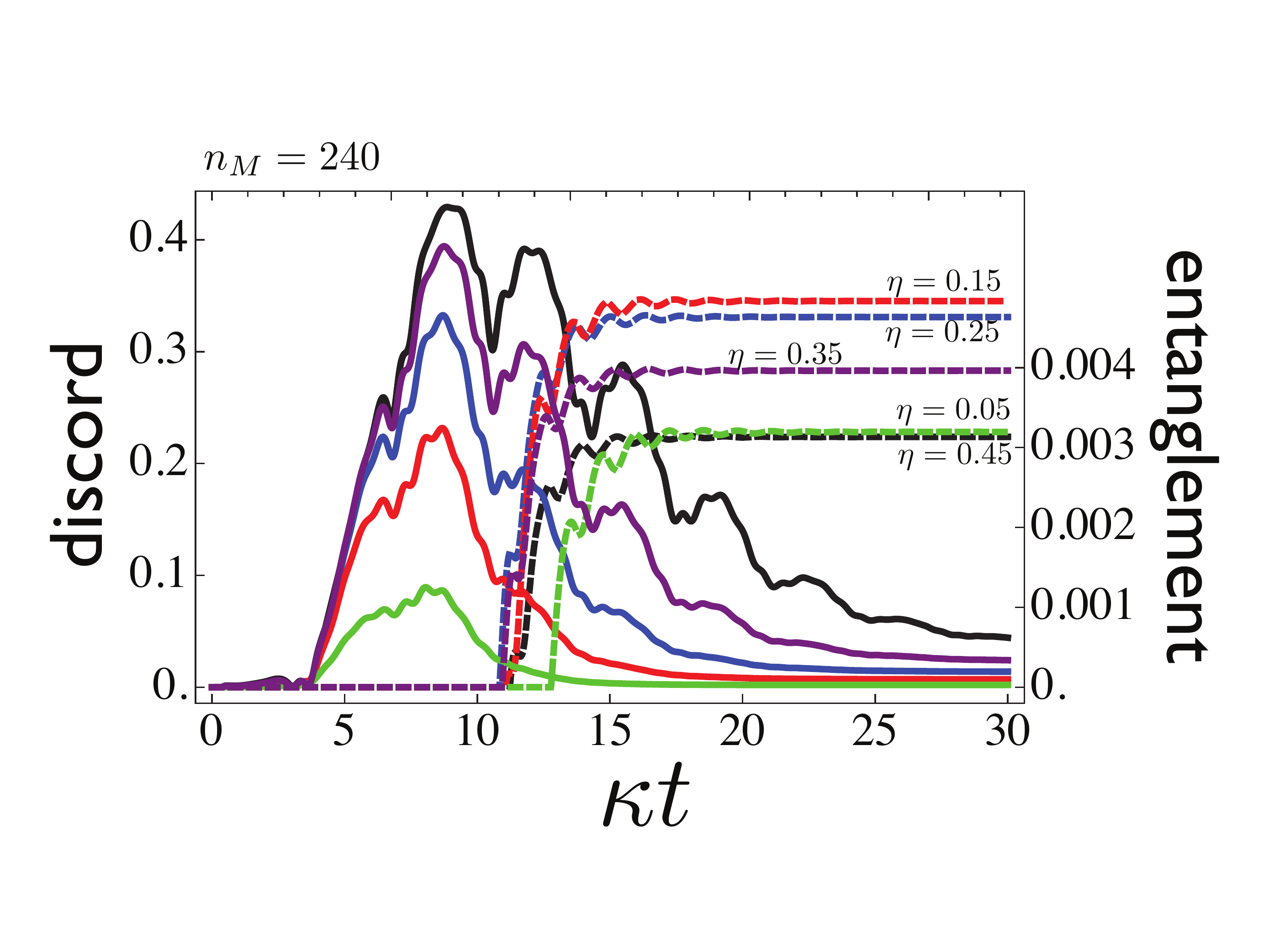}
		\caption{(Color online) Time evolution of the Gaussian discord ${\cal D}_G$ between $O_1$ and $O_2$ (continuous lines) and of the logarithmic negativity $\mathcal E$ 
		across the $O_1 O_2/M_1 M_2$ bipartition (dashed lines). The various color refers to different values of beam splitter
		transmissivity $\eta$.   
 We have set $n_M=240$ and taken OMs  to be initially in the vacuum state. Time is measured in units of $\kappa^{-1}$.}		\label{Figsupp2}
	\end{center}
\end{figure}

The dependence of ${\cal D}_G$ and ${\cal E}$ upon the transmissivity $\eta$ of the optical fiber is analyzed in Fig.~\ref{Figsupp2}. Again, for all the values of $\eta$ we have tested we observe a temporal evolution which is consistent with the one reported in Fig. 5 of the main text. 
We notice also that,  analogously to what seen for the simplified model of Fig. 2, the Gaussian discord tends to increase with
$\eta$. On the contrary ${\cal E}$ exhibits a non monotonic
behavior in $\eta$. Interestingly, for large values of $\eta$,  ${\cal E}$ and ${\cal D}_G$ exhibits oscillations which are probably
 associated with multiple reflections of the transmitted signals.

\begin{figure}[ht]
	\begin{center}
	\includegraphics[trim=0pt 0pt 0pt 0pt, clip, width=0.4\textwidth]{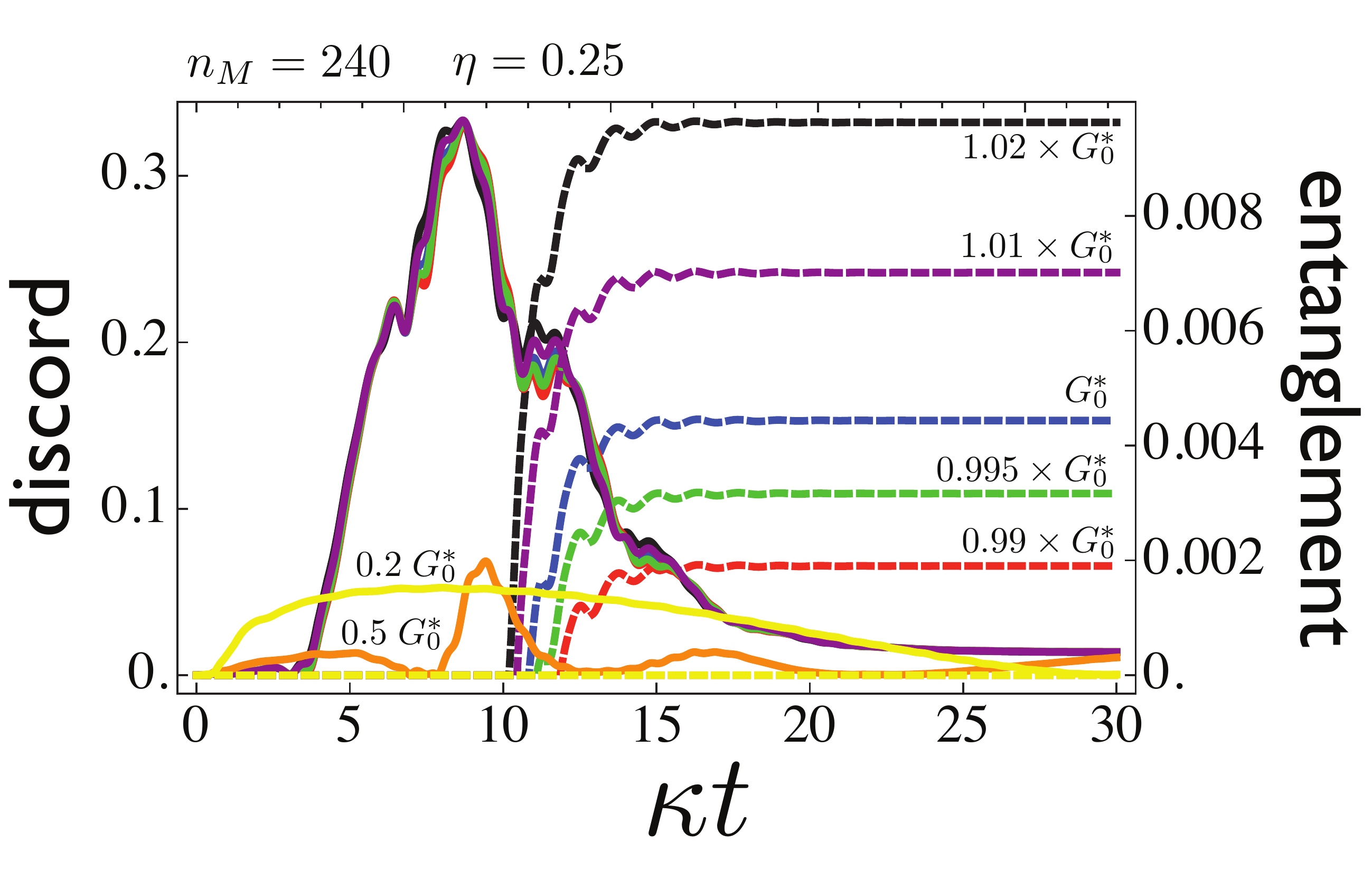}
\caption{(Color online) Time evolution of the Gaussian discord ${\cal D}_G$ between $O_1$ and $O_2$ (continuous lines) and of the logarithmic negativity $\mathcal E$ 
		across the $O_1 O_2/M_1 M_2$ bipartition (dashed lines). The various color refers to different values of the effective
		opto-mechanical coupling constant $G_0$ (here $G_0^*$ stands for the value used in the main text, i.e. 24 Hz).  
 We have set $\eta=0.25$, $n_M=240$ and taken OMs  to be initially in the vacuum state. Time is measured in units of $\kappa^{-1}$.}
		\label{Figsupp3}
	\end{center}
\end{figure}

Finally in Fig.~\ref{Figsupp3} we report the time dependence
of ${\cal D}_G$ and ${\cal E}$ for different values of the opto-mechanical coupling. As long as the latter is sufficient large, 
${\cal D}_G$  is not affected by variation of this parameter. In the discord production in fact $G_0$ enters only indirectly as the
mechanisms that transfers the thermal excitation from the
mechanical oscillator the optical modes. Only when the opto-mecahnical is sufficiently small (orange and yellow curves), ${\cal D}_G$ gets significantly reduced.
On the contrary, ${\cal E}$ is directly affected by $G_0$: a small
decrease in this parameter results in a strong reduction of the
resulting entanglement level. 


%
%
%
%
%
%

\newpage
\mbox{}


\begin{thebibliography}{50}
\bibitem{NOTA1}
The case $\eta\ug1$ is critical since, for such value, the normal mode whose associated annihilation operator is $\hat{a}_1\meno\hat{a}_2$ becomes decoupled from the environment and evolves unitarily. Therefore, the system final state depends on the initial conditions and is not uniquely determined by the dissipative dynamics. Yet, a realistic setup will always have additional noise sources which, no matter how small, prevent this behavior from occuring.
\bibitem{NOTANEW1} Although the optical baths are at zero temperature, discord can still be created because now additional MMs, alongside their respective baths, interact with the OMs.
\end{thebibliography}
\end{document}